\documentclass[%
reprint,
 amsmath,amssymb,
 aps,
]{revtex4-2}

\usepackage{graphicx}% Include figure files
\usepackage{dcolumn}% Align table columns on decimal point
\usepackage{bm}% bold math
\usepackage{color}
\usepackage{soul}
\newcommand{\red}{\textcolor{black}} % revised parts in red

\begin{document}

\title{Testing Modified Gravity Theories with Numerical Solutions of the External Field Effect in Rotationally Supported Galaxies}

\author{Kyu-Hyun Chae}
\affiliation{Department of Physics and Astronomy, Sejong University, 209 Neungdong-ro Gwangjin-gu, Seoul 05006, Republic of Korea}
\email{chae@sejong.ac.kr, kyuhyunchae@gmail.com}

\author{Federico Lelli}
\affiliation{INAF, Arcetri Astrophysical Observatory, Largo Enrico Fermi 5, I-50125 Florence, Italy} 
%\email{FL: federico.lelli@inaf.it}

\author{Harry Desmond}
\affiliation{McWilliams Center for Cosmology, Department of Physics, Carnegie Mellon University, 5000 Forbes Ave, Pittsburgh, PA 15213, USA}
\affiliation{Institute of Cosmology \& Gravitation, University of Portsmouth, Dennis Sciama Building, Portsmouth, PO1 3FX, UK}
%\email{HD: hdesmond@andrew.cmu.edu}

\author{Stacy S. McGaugh}
\affiliation{Department of Astronomy, Case Western Reserve University, Cleveland, OH 44106, USA} 
%\email{SSM: ssm69@case.edu}

\author{James M. Schombert}
\affiliation{Department of Physics, University of Oregon, Eugene, OR 97403, USA}
%\email{JMS: schombe@gmail.com}

\begin{abstract}

The strong equivalence principle is violated by gravity theories of Milgromian dynamics (MOND) through the action of the external field effect. We test two different Lagrangian theories AQUAL and QUMOND based on their numerical solutions of the external field effect, by comparing two independent estimates of the mean external field strength of the nearby universe: a theory-deduced value from fitting the outer rotation curves of 114 galaxies and an empirical value from the large-scale distribution of cosmic baryons. The AQUAL-deduced external field strength from rotation curves agrees with that from the large-scale cosmic environment, while QUMOND-deduced value is somewhat higher. This suggests that AQUAL is likely to be preferred over QUMOND as an effective non-relativistic limit of a potential relativistic modified gravity theory.

\end{abstract}

%\keywords{Non-standard theories of gravity; Modified Newtonian dynamics; Disk galaxies}

\maketitle

\section{Introduction}  \label{sec:intro}

The nearly-flat rotation curves (RCs) of spiral galaxies discovered in the 1970s \citep{Rubin1970,Rubin1978,Bosma1978} have been interpreted as evidence for particle dark matter. Alternatively, this characteristic feature of RCs may imply the departure from standard dynamics at accelerations below a critical value $a_0\approx 1.2\times 10^{-10}$~m~s$^{-2}$, as was first proposed by M. Milgrom \cite{Mil1983} in a general theoretical framework called modified Newtonian dynamics (MOND).   

The MOND hypothesis led to the construction of specific non-relativistic Lagrangian theories of gravity, such as the Aquadratic-Lagrangian (AQUAL) theory \citep{BM1984} and the Quasilinear MOND (QUMOND) \citep{Mil2010} theory. Relativistic theories of MOND have also been under active development \citep{Mil2009,SZ2021}. The current status of MOND research is reviewed in \cite{FM2012}, \cite{Mil2014}, \cite{merritt2020}, and \cite{BZ2021}. 

A unique feature of MOND is the external field effect (EFE), experienced by test particles in a self-gravitating system falling freely under a constant external field.  The $\Lambda$ cold dark matter ($\Lambda$CDM) cosmological model invokes dark matter and dark energy as a consequence of retaining general relativity and hence the strong equivalence principle (SEP). MOND gravity violates the SEP through the EFE. The EFE therefore provides a strong test of MOND \citep{MM2013}, provided dark matter is not fortuitously distributed \textit{just so} to imitate this effect \citep{Oria2021}. 

In rotationally supported disk galaxies, the EFE causes the nearly-flat RC to decline at an acceleration typically much weaker than $a_0$. Weakly declining RCs in the outskirts of disk galaxies have been reported \citep{Haghi2016,Hees2016,Chae2020b,Chae2021}. The works of Chae et al. \cite{Chae2020b, Chae2021}, hereafter Paper I \& II, have demonstrated that an unbiased sample of $>150$ RCs exhibit a decline in an average sense and that RCs in higher density environments are more likely to decline than those in voids, consistent with the generic MOND prediction for the EFE. These works used a toy model for the EFE based on a one-dimensional approximation \citep{FM2012} because numerical solutions were not available at that time. 

\begin{figure}
  \centering
  \includegraphics[width=0.9\linewidth]{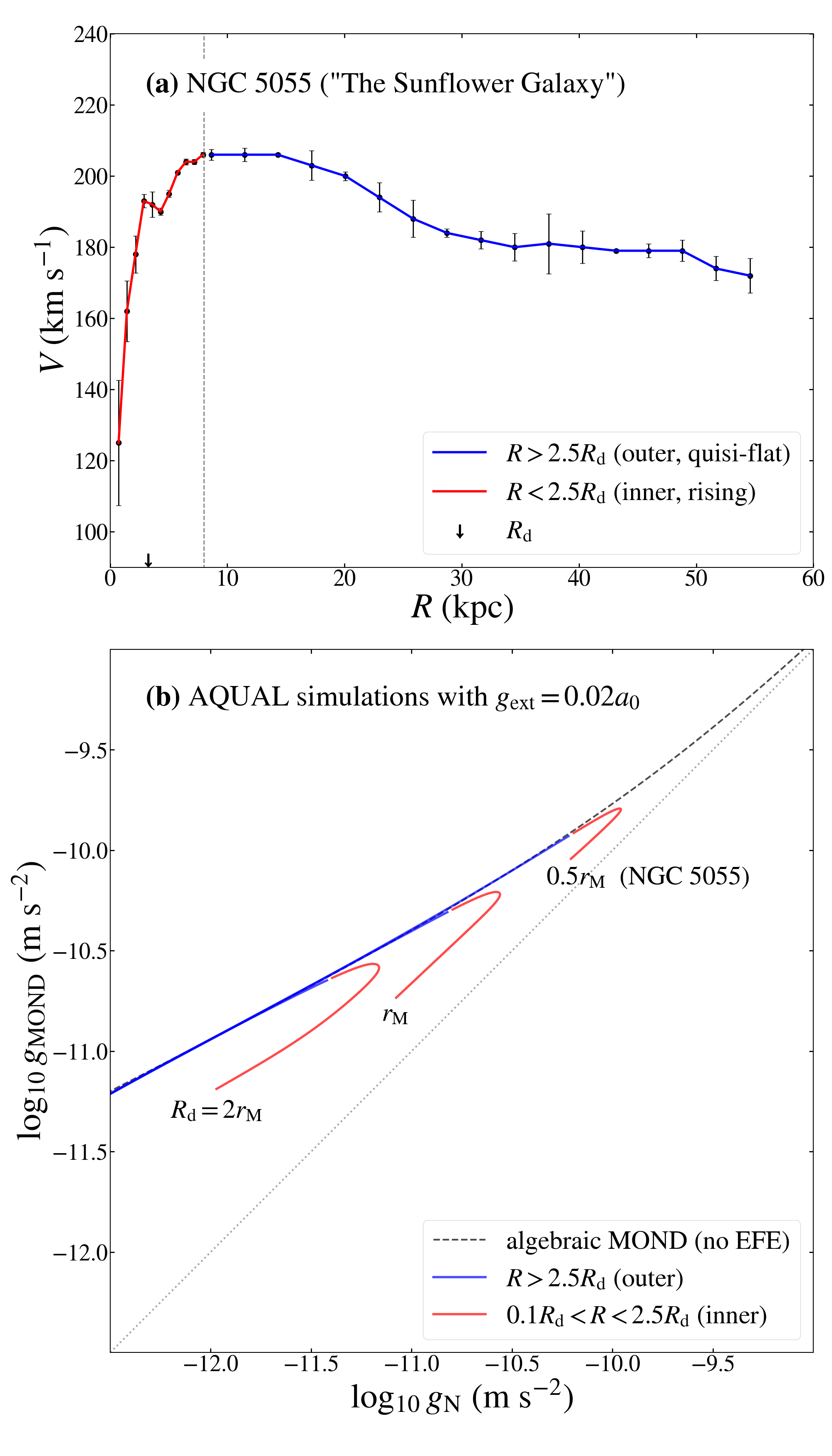}
    \vspace{-0.5truecm}
    \caption{\small 
    \textbf{(a)} The observed rotation curve of a galaxy (NGC\,5055) with a well-defined outer part (blue). The inner rising part corresponds to $R<2.5R_{\rm d}$, where $R_{\rm d}$ is the disk scale radius (Eq.\,\ref{eq:expdisk}). \textbf{(b)} The radial acceleration relation from AQUAL numerical simulations of a disk under a weak external field $g_{\rm ext}=0.02a_0$ for $a_0=1.2\times 10^{-10}$m~s$^{-2}$. Different lines show exponential disks with the same mass $M$ but different scale radius $R_{\rm d}=0.5r_{\rm M}$ (corresponding approximately to NGC\,5055), $r_{\rm M}$, and $2r_{\rm M}$, where $r_{\rm M} \equiv \sqrt{GM/a_0}$ is the MOND radius. The inner and outer parts are indicated by red and blue colors as in panel \textbf{(a)}. The inner parts deviate from the algebraic MOND relation (black dashed curve)  for non-spherical mass distributions \citep{BM1995}, even when the external field is very weak \citep{CM2021}, thus they are excluded in our EFE analyses.
    } 
   \label{inner}
\end{figure}

Here we use numerical solutions \citep{CM2021} of the two nonlinear theories AQUAL and QUMOND to test and compare these theories through the EFE. For this we consider 114 rotationally supported galaxies, whose RCs have well-defined outer parts (see Fig.\,\ref{inner}(a) for an example \red{and~\citep{Chae2022} for further details}), from the Spitzer Photometry and Accurate Rotation Curves (SPARC) database \citep{Lel2016}. Thus, we deduce for the first time AQUAL- and QUMOND-based values of the mean Newtonian field of the nearby universe to compare with the value estimated directly from the large-scale structure of baryons (Paper~II). Numerical values of accelerations are given in units of $10^{-10}$~m~s$^{-2}$ unless specified otherwise.

\section{Theory}  \label{sec:theory}

The basic tenet of MOND for an isolated gravitational system is encapsulated in the following algebraic relation between the Newtonian gravitational field $\mathbf{g}_{\rm N}$ from a given baryonic mass distribution and the kinematic acceleration $\mathbf{g}=d^{2}\mathbf{q}/dt^{2}$ for position $\mathbf{q}$:

\begin{equation}
 \mu(g/a_0)\mathbf{g}=\mathbf{g}_{\rm N},~~ {\rm or}~~ \mathbf{g}=\nu(g_{\rm N}/a_0)\mathbf{g}_{\rm N},
    \label{eq:algemond}
\end{equation}
where $\mu(X)$ and $\nu(Y)$ are interpolating functions (IFs). For $X\equiv g/a_0$ and $Y\equiv g_{\rm N}/a_0$, the IFs have the following asymptotic behavior: $\mu(X\gg 1)\approx 1$, $\mu(X\ll 1)\approx X$, $\nu(Y\gg 1)\approx 1$, and $\nu(Y\ll 1)\approx Y^{-1/2}$. In this work we use the `simple' IF \citep{FB2005} of $\mu(X)=X/(1+X)$ and $\nu(Y)=1/2+\sqrt{1/4+1/Y}$ which describes well galaxy data at $g_{\rm N}< 10^{-8}$~m~s$^{-2}$ \citep{Chae2019}.

Eq.\,(\ref{eq:algemond}) is often used as a convenient representation of MOND but its general validity is limited. In modified gravity theories Eq.\,(\ref{eq:algemond}) is strictly correct only for orbits in spherical mass distributions \citep{BM1984, Mil2010}, while in modified inertia theories it applies only to circular orbits in any mass distribution \citep{Mil1994}. If the 3-dimensional nature of galaxies is not considered, Eq.\,(\ref{eq:algemond}) allows an analytic description of the EFE on the internal acceleration $\mathbf{g}$ by replacing $\mu$ or $\nu$ with an appropriate function of the external field strength \citep{FM2012}. This is the approach adopted in previous analyses (Paper I, II).

The AQUAL field equation for a mass distribution $\rho$ is given by 
\begin{equation}
  \mathbf{\nabla} \cdot \left[ \mu\left(|\mathbf{\nabla}\Phi|/a_0\right)\mathbf{\nabla}\Phi \right] = 4\pi G\rho,
 \label{eq:aqual}
\end{equation}
where $\Phi$ is the MOND potential, while the QUMOND field equation is given by
\begin{equation}
  \mathbf{\nabla}^2 \Phi=\mathbf{\nabla}\cdot \left[ \nu\left( |\mathbf{\nabla}\Phi_{\rm N}|/a_0 \right) \mathbf{\nabla}\Phi_{\rm N} \right],
    \label{eq:qumond}
\end{equation}
where $\Phi_{\rm N}$ is the Newtonian potential sourced by $\rho$, i.e.\ $\mathbf{\nabla}^2 \Phi_{\rm N} = 4\pi G \rho$. Here $G$ is Newton's gravitational constant and $a_0\approx 1.2\times 10^{-10}$~m~s$^{-2}$ \citep{McGaugh2012,MLS2016}. When an external field is present, the total potential $\Phi$ in Eq.\,(\ref{eq:aqual}) satisfies the boundary condition $-\mathbf\nabla\Phi\rightarrow \mathbf{g}_{\rm ext}$ (the external MOND field) and the internal acceleration is given by $-\mathbf{\nabla}\Phi-\mathbf{g}_{\rm ext}$. In Eq.\,(\ref{eq:qumond}), we obtain the internal acceleration after the application of $\mathbf{\nabla}\Phi_{\rm N} \rightarrow \mathbf{\nabla}\Phi_{\rm N}-\mathbf{g}_{\rm N,ext}$ where $\mathbf{g}_{\rm N,ext}$ is the external Newtonian field (from solving the standard Poisson's equation for all external baryonic mass). Throughout we use the notations $e \equiv {g}_{\rm ext}/a_0$, $e_{\rm N}\equiv {g}_{\rm N,ext}/a_0$, and $\tilde{e} \equiv \sqrt{e_{\rm N}}$, and we use the relation $\tilde{e}=e/\sqrt{1+e}$ assuming the simple IF.

Eqs (\ref{eq:aqual}) and (\ref{eq:qumond}) are non-linear and can only be solved numerically. For disk systems, one interesting prediction of these modified gravity theories is that in the inner part (within about two disk scale lengths) the centripetal acceleration deviates downward from the algebraic MOND relation (Eq.\,\ref{eq:algemond}), whether an external field is present or not \citep{BM1995, CM2021}. Fig.\,\ref{inner}(b) shows examples for simulated disks under $e=0.02$. The surface density projected along the symmetry axis of the disk is given by the exponential profile 
\begin{equation}
    \Sigma(R)=\Sigma_0\exp(-R/R_{\rm d}), 
    \label{eq:expdisk}
\end{equation}
where $R_{\rm d}$ is the disk scale radius. For such a weak external field, little deviation is expected in the outer part of the RCs within the acceleration range probed by SPARC galaxies ($g_{\rm N} > 10^{-12.5}$~m~s$^{-2}$). However, sizable deviations unrelated to the EFE are expected in the inner part due to the non-spherical symmetry of disk galaxies (no such deviations are expected for circular orbits in modified inertia theories). A strong EFE further adds to the complexity of the inner part. Thus, we will not consider the inner rising part in this study. 

Numerical studies of disk galaxies in AQUAL \citep{CM2021} and QUMOND \citep{CM2021, Zon2021,Oria2021} have obtained radial acceleration relations (RARs) between $g_{\rm N}$ and $g_{\rm MOND}$ depending on the external field strength for a large acceleration range, except for the inner rising part. The AQUAL EFE-dependent RAR is given by \cite{CM2021} as
\begin{equation}
      g_\text{AQUAL} = g_{\rm N} \nu(y_\beta) \left[1+\tanh\left(\frac{\beta e_{\rm N}}{g_{\rm N}/a_0}\right)^{\gamma} \frac{\hat{\nu}(y_\beta)}{3} \right],
    \label{eq:aqualfit}
\end{equation}
where $y_\beta \equiv \sqrt{(g_{\rm N}/a_0)^2+(\beta e_{\rm N})^2}$,  $\beta=1.1$, $\gamma=1.2$ and $\hat{\nu}(y)\equiv d \ln\nu(y)/d\ln y$. Eq.\,(\ref{eq:aqualfit}) is the azimuthally averaged quantity on a plane, so the dependence on the orientation of the external field is minor \citep{CM2021}. The QUMOND EFE-dependent RAR is given by \cite{Zon2021} as 
\begin{equation}
     g_\text{QUMOND} = g_{\rm N} \nu(y_1) \left[1+\tanh\left(\frac{0.825e_{\rm N}}{g_{\rm N}/a_0}\right)^{3.7} \frac{\hat{\nu}(y_1)}{3} \right],
    \label{eq:qumondfit}
\end{equation}
where $y_1 = \sqrt{(g_{\rm N}/a_0)^2+e_{\rm N}^2}$ (see also \cite{Oria2021,CM2021}). 

In a disk galaxy, the internal Newtonian gravitational field can be computed from the observed distribution of baryons (gas and stars) and is usually indicated as $g_{\rm bar}$ \citep{MLS2016,Lelli2017}. The radial (centripetal) acceleration is measured as $g_{\rm obs}=V^2/R$, where $V$ is the average rotation speed in a ring centered at radius $R$. Hence $g_{\rm obs}$ corresponds to the azimuthally averaged radial acceleration in the disk midplane and can be matched with the dynamical acceleration of Eqs~(\ref{eq:aqualfit}) or (\ref{eq:qumondfit}).

\section{Data Analysis}  \label{sec:analysis}

We consider a sample of 162 SPARC galaxies (Paper II) excluding only 13 galaxies with low quality rotation curves ($Q=3$) from the SPARC database \citep{Lel2016}. This provides 3200 measurements of $g_{\rm bar}$ and $g_{\rm obs}$. Given that Eqs (\ref{eq:aqualfit}) and (\ref{eq:qumondfit}) apply only to the outer RCs (see Fig.\,\ref{inner}), we exclude 1479 data points from the inner rising part and are left with 1721 data points from 114 galaxies (this means that 48 out of 162 galaxies lack measured outer parts). The median inner-outer transition radius for all 175 SPARC galaxies is $2.5 R_{\rm d}$ or $1.7 R_{\rm eff}$ where $R_{\rm d}$ and $R_{\rm eff}$ are the exponential disk scale length and effective (i.e.\ half-mass) radius \citep{Lel2016}. 

\red{We use two complementary approaches.} One is a statistical approach where $x\equiv\log_{10}g_{\rm bar}$ and $y\equiv\log_{10}g_{\rm obs}$ from different galaxies are stacked and modeled with a common $\tilde{e}$, which represents the mean gravitational field of the nearby Universe.  We perform a joint fit of $\tilde{e}$ and $a_0$; this is important because the numerical value of $a_0$ has usually been derived neglecting the EFE. We adopt a Bayesian approach and use a Gaussian likelihood function:
\begin{equation}
    \ln\mathcal{L} =-\frac{1}{2} \sum_i \left( \frac{\Delta_{\bot,i}^2}{s_\theta^2\sigma_{x_i}^2 +c_\theta^2\sigma_{y_i}^2} + \ln[2\pi(s_\theta^2\sigma_{x_i}^2 +c_\theta^2\sigma_{y_i}^2 )] \right),
    \label{eq:likelihood}
\end{equation}
where $\Delta_{\bot,i}$ is the orthogonal distance of point $(x_i, y_i)$ from the model curve and its error is contributed by \red{$s_\theta\sigma_{x_i}$ and $c_\theta\sigma_{y_i}$ with $s_\theta\equiv\sin\theta$ and $c_\theta\equiv\cos\theta$ for the angle $\theta$ of the tangent line of the curve from the $x$-axis.} The use of orthogonal distances guarantees a robust fit to the data with uncertainties in both ($x$ and $y$) directions as we have verified with simulated data. From a Markov chain Monte Carlo (MCMC) procedure, posterior probability distribution functions (PDFs) of $\tilde{e}$ and $a_0$ are derived. All MCMC simulations are carried out using the public package {\tt emcee} \citep{emcee}. A flat prior on $\tilde{e}$ is set between $-2.6<\log_{10}\tilde{e}<-0.15$ using the limits derived in Paper~II from the large-scale structure of cosmic baryons. A Gaussian prior is set on $a_0=(1.24\pm 0.14)\times 10^{-10}$~m~s$^{-2}$ based on the baryonic Tully-Fisher relation of gas-dominated galaxies \citep{McGaugh2012}, obtained largely independent of SPARC data. Because the Gaussian width is broad, this prior is practically similar to a flat prior.

We estimate $g_{\rm bar}$ and $g_{\rm obs}$ using empirical quantities taken from the SPARC database. They depend on the measured disk inclination $i$, galaxy distance $D$, mass-to-light ratios of stellar disk ($\Upsilon_{\rm disk}$) and bulge ($\Upsilon_{\rm bulge}$), and gas-to-HI mass ratio ($\Upsilon_{\rm gas}$). We take $\Upsilon_{\rm disk}=0.5,\;\Upsilon_{\rm bulge}=0.7\;\mathrm{M}_\odot/\mathrm{L}_{\odot}$, and $\Upsilon_{\rm gas}=1.33$ correcting atomic gas for the presence of primordial helium \citep{MLS2016} along with common uncertainties of 25\% for $\Upsilon_{\rm disk}$ and $\Upsilon_{\rm bulge}$ and 20\% for $\Upsilon_{\rm gas}$.

\red{The advantage of this statistical approach is that all outer data points from all galaxies can be used on an equal footing. The caveat is that individual peculiarities are ignored. In reality, mass-to-light ratios of the disk and the bulge may vary from galaxy to galaxy. More importantly, galaxies are under different environments so that each galaxy should have its own value of $\tilde{e}$ under the MOND framework. Our working assumption it that these peculiarities are averaged out by stacking $>100$ galaxies (see~\cite{Chae2022} for further details).}

In the other approach, we carry out Bayesian fits to individual RCs to infer PDFs of  galaxy parameters \{$i$, $D$, $\Upsilon_{\rm disk}$, ($\Upsilon_{\rm bulge}$,) $\Upsilon_{\rm gas}$\}, together with critical acceleration $a_0$ and individual external field strength $\tilde{e}$, which constitute 6 (or 7) free parameters. Priors on galactic parameters are the same as in Paper~I. 

The advantage of this approach is that galaxy-specific parameters including mass-to-light ratios and their uncertainties are derived. Also, $a_0$ and $\tilde{e}$ may be well-constrained for exceptional systems with good quality RCs. However, because the reported uncertainties of individual RCs vary from galaxy to galaxy as the sample is a collection from various heterogeneous observations, the Bayesian inferred individual values of $\tilde{e}$ and their uncertainties may, in general, need to be taken with caution.

\begin{figure*}
 \hspace{-0.29truecm}  
  \centering
  \includegraphics[width=1.012\linewidth]{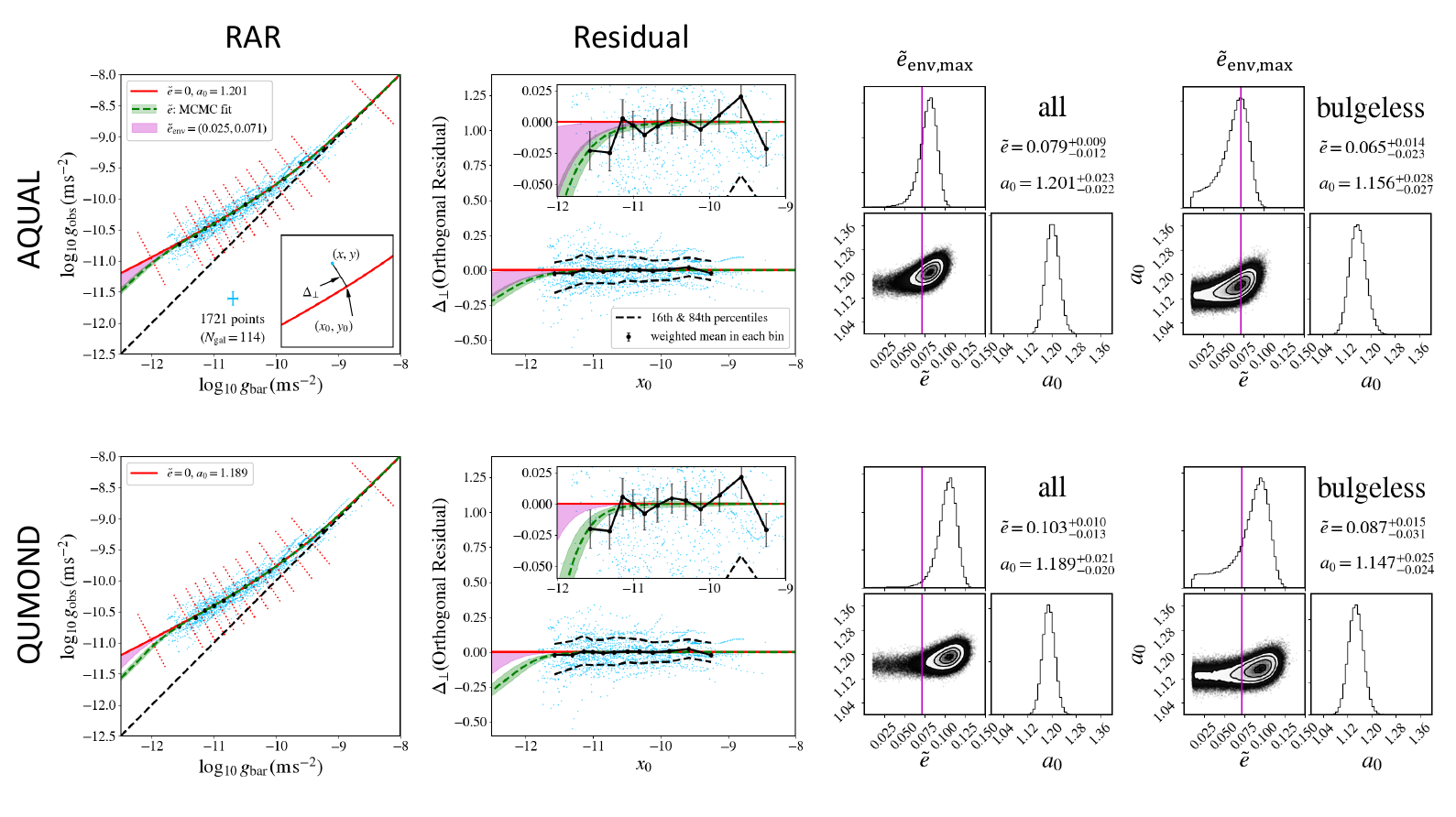}
    %\vspace{-0.5truecm}  
    \caption{\small 
    \textbf{Fitting results from the statistical approach}. Top and bottom rows: AQUAL and QUMOND fits of the stacked SPARC data. \emph{1st column:} Data points (cyan dots) are compared with the EFE-free algebraic MOND relation (red line). Green dashed curves with a band show the result of the MCMC joint fit of the mean external field $\tilde{e}$ and $a_0$ taking into account individual error bars (typical ones are indicated in the bottom-right corner of the panels). The magenta band represents the possible range of the mean environmental field $\tilde{e}_\text{env}$ from cosmic baryons. Red dotted lines define bins orthogonal to the red solid curve. The inset shows the definition of $x_0$ and $\Delta_\bot$ (orthogonal residual from the red solid curve).  \emph{2nd column:} $\Delta_\bot$ as a function of $x_0$. The inset shows \red{weighted means} and their uncertainties in the bins. \emph{3rd column:} Corner plots showing the posterior PDFs of $\tilde{e}$ and $a_0$. The purple vertical lines indicate the maximum value of $\tilde{e}_\text{env}$ from Paper II. \emph{4th column:} Same as the 3rd column but for a subsample of galaxies without bulges.
    } 
   \label{efeaqual}
\end{figure*}

Given that the maximum number of free parameters in the Bayesian fits is 7 as specified above, we consider only galaxies possessing at least 8 rotation velocities from the outer RCs. By this criterion, 73 out of 162 galaxies are selected. From modeling these galaxies, we notice that RCs covering a narrow acceleration range cannot constrain well the fitting functions. Thus, we further apply a dynamic range cut $(\Delta x_0)_{1/2} \equiv |x_{\rm 0,outermost} - x_{\rm 0,median}| > 0.2$ so that the weaker acceleration part of the RC covers at least twice the typical uncertainty of $x$. Here $x_0$ for a given $x$ is defined in Fig.\,\ref{efeaqual}. By this cut, we are left with 65 galaxies that can be used for individual Bayesian modeling. 

\section{Results}  \label{sec:result}

Fig.\,\ref{efeaqual} summarizes our fitting results for AQUAL and QUMOND based on the statistical approach. We present results not only for all selected galaxies but also for a subsample of galaxies without bulges. Due to significant non-circular motions in the bulge, the reported rotation velocities and their uncertainties may be less reliable in the inner RC if a bulge is present. Thus, it is interesting to consider a sample excluding those with bulges.

For AQUAL, we have \red{$\tilde{e}=0.079_{-0.012}^{+0.009}$} (68\% confidence limit, hereafter) based on all galaxies, or \red{$\tilde{e}=0.065_{-0.023}^{+0.014}$} based on bulgeless galaxies. \red{The subsample of bulgeless galaxies is considered to investigate possible systematic biases when a bulge is present.} Two values agree with each other and overlap well with the range $\tilde{e}_{\rm env,min}(=0.025\pm 0.001)\le \tilde{e}_{\rm env} \le \tilde{e}_{\rm env,max}(=0.071\pm 0.001)$ for the mean environmental field in the nearby universe from Paper~II, where $\tilde{e}_{\rm env,max}$ ($\tilde{e}_{\rm env,min}$) refers to the limiting value when intergalactic (and circumgalactic) baryons (those not in galaxies and clusters of galaxies) make maximal (no) contribution to $\tilde{e}_{\rm env}$. \red{(The values of $\tilde{e}_{\rm env,max}$ and $\tilde{e}_{\rm env,min}$ are unchanged even if only bulgeless galaxies are used.)} The above results indicate that the AQUAL-deduced values of $\tilde{e}$ prefer a high value close to $\tilde{e}_{\rm env,max}$.

Posterior PDFs of $\tilde{e}$ are not normal but have tails towards zero as shown in the 3rd and 4th columns of Fig.\,\ref{efeaqual}. Table\,\ref{tab:prob} shows probabilities of $p(\tilde{e}<\tilde{e}_{\rm env,max})$ and $p(\tilde{e}<\tilde{e}_{\rm env,min})$. For AQUAL, the probability of $p(\tilde{e}<\tilde{e}_{\rm env,max})$ is sufficiently high for both the full and bulgeless samples. \red{Table\,\ref{tab:prob} also shows an intermediate case for the mean probability $\tilde{e}_{\rm env,mean}\equiv(\tilde{e}_{\rm env,max}+\tilde{e}_{\rm env,min})/2$ of the two boundaries. The probability of $0.21$ for the bulgeless sample is sufficiently high while $0.02$ for the full sample is not. This may indicate that AQUAL would be consistent only with an environmental field higher than the mean of the two extremes or data of galaxies with bulges need to be taken with caution. }

For QUMOND, the fitted values are \red{$\tilde{e}=0.103_{-0.013}^{+0.010}$} based on all galaxies, or \red{$\tilde{e}=0.087_{-0.031}^{+0.015}$} based on bulgeless galaxies. These values are higher than the AQUAL values by $\approx 0.023$ and less consistent with the environmental field. \red{Probability tests given in Table\,\ref{tab:prob} also show that QUMOND is less consistent with the environmental field.}

 \begin{table*}
\caption{Probability tests. The third and \red{fourth} columns give the probability for the external field strength derived from rotation curves to be less than the cosmic environmental field from the baryonic large-scale structure, when intergalactic baryons are maximally ($\tilde{e}_{\rm env,max}$) or minimally ($\tilde{e}_{\rm env,min}$) clustered. \red{The last column is for the mean of the two limits, $\tilde{e}_{\rm env,mean}\equiv(\tilde{e}_{\rm env,max}+\tilde{e}_{\rm env,min})/2$}. }\label{tab:prob}
\begin{center}
  \begin{tabular}{llccc}
  \hline
 Model & sample & $p(\tilde{e}<\tilde{e}_{\rm env,max})$  & $p(\tilde{e}<\tilde{e}_{\rm env,min})$  & $p(\tilde{e}<\tilde{e}_{\rm env,mean})$\\
 \hline
AQUAL & all        & 0.25   & 0.004  & 0.02     \\ 
AQUAL & bulgeless  & 0.68    & 0.06  & 0.21     \\ 
QUMOND & all       &  0.04  &  0.004 & 0.01   \\ 
QUMOND & bulgeless & 0.27    & 0.04 & 0.12      \\ 
  \hline
\end{tabular}
\end{center}
\end{table*}

Individual Bayesian modeling provides individually fitted values of $\tilde{e}$ and $a_0$ for 65 galaxies. The value of $\tilde{e}$ can usually be poorly constrained in individual galaxies (Paper I, II), but it is important to check whether individual values of $\tilde{e}$ are \emph{statistically} consistent with their environmental estimate within the errors. We define the quantity $d\equiv (u - v)/\sqrt{\sigma_u^2+\sigma_v^2}$ where $u\equiv\log_{10}\tilde{e}_{\rm env,max}$, $v\equiv\log_{10}\tilde{e}$, and $\sigma_u$ and $\sigma_v$ are their estimated uncertainties. Thirty six out of 65 galaxies are in the Sloan area and have $\tilde{e}_{\rm env,max}$ values from Paper~II. For this subsample, we find $\langle d\rangle = 0.35\pm 0.31$ for AQUAL models and $\langle d\rangle = 0.57\pm 0.34$ for QUMOND models. The AQUAL results indicate good galaxy-by-galaxy agreement between the independent external field estimates from the RC fits and the large-scale distribution of baryons in agreement with the statistical fit results of the stacked data shown in Fig.\,\ref{efeaqual}.

\section{Discussions and Conclusion} \label{sec:disc}

We have addressed the question of whether numerical solutions of Lagrangian theories of modified gravity (AQUAL and QUMOND) give an RC-fitted external field strength consistent with that from the large-scale distribution of baryons from Paper~II.

The baryons that reside in galaxies and clusters of galaxies account for about one eighth of the total \citep{Shull2012} from Big Bang nucleosynthesis (which is expected to proceed normally in MOND \cite{Sanders1998}), with the remainder being in the intergalactic (and circumgalactic) media \citep{Nicastro2018,Macquart2020,Schaan2021}. The distribution of these intergalactic baryons is uncertain, so we have considered bracketing limits in which these baryons are completely uniform in distribution (so provide no enhancement to the EFE) or maximally correlated with observed galaxies.

Our investigation reveals two important results. First, the AQUAL-deduced field overlaps well with the environmental range while the QUMOND-deduced field does not as well. Thus, AQUAL is preferred, though QUMOND is not excluded given the uncertainties in the galaxy data. \red{The consistency between the theory(in particular, AQUAL)-deduced value and the environmental field is not a trivial result. A small systematic change in outer RCs can easily lead to an order of magnitude discrepancy in $\tilde{e}$. }

Second, the preferred environmental field strength is close to the maximum value. Thus, MONDian gravity implies that the spatial distribution of intergalactic baryons is correlated with the large scale structure of galaxies, as expected in structure formation with MOND \citep{LKZ2008}.

\red{Given that the outer parts of galactic rotation curves can be successfully described by numerical solutions of EFE in AQUAL as shown in this work, there naturally arises the question of whether AQUAL could predict correctly the inner parts (see Fig.\,\ref{inner}). This is indeed the case as demonstrated in \cite{Chae2022}.}

\red{Although AQUAL appears to perform better than QUMOND in the EFE phenomenology, the two theories are similar in many aspects of galactic dynamics. For example, as shown in \cite{CM2021}, the predictions of the two theories on the inner parts of flattened systems are very similar. QUMOND has the advantage of being more tractable mathematically and numerically. Thus, the theory may well continue to be used in various studies such as numerical simulations of galactic dynamics (e.g.\ \cite{Banik2020}) although its prediction on EFE may not be as accurate as AQUAL. It may also be possible to use an effective external field in QUMOND rather than the true external field to compensate for the small difference with AQUAL.}

\red{In conclusion, the AQUAL theory of MOND provides a successful description of galactic rotation curves, and thus AQUAL may be preferred over QUMOND as an effective non-relativistic limit to a potential relativistic theory of MOND (or other modified) gravity. }

\section*{Acknowledgments}
 This manuscript benefited from comments from anonymous referees. This research was supported by the National Research Foundation of Korea (NRF) grant funded by the Korea government(MSIT) (No.\ NRF-2022R1A2C1092306). HD was supported by a McWilliams Fellowship and a Royal Society University Research Fellowship (grant no. 211046).

\end{document}